\newcommand{\lya}{Ly$\alpha$ }
\newcommand{\lyaf}{Ly$\alpha$ forest}

\def\h2{${\rm\,H_2}$}

\makeatletter 

\def\vol#1  {{{#1}{\rm,}\ }}
\def\lya{{\rm Ly}\alpha}

\newcount\refno
\newcount\rfno
\def\eq{$^{\the\refno\ }$\advance\refno by 1}
\def\ad{\advance\rfno by 1}

\def\clock{\count0=\time \divide\count0 by 60
     \count1=\count0 \multiply\count1 by -60 \advance\count1 by \time
     \number\count0:\ifnum\count1<10{0\number\count1}\else\number\count1\fi}

\def\myputfigure#1#2#3#4#5%
{\hskip0.03\textwidth\vskip#5pt%\hfill
\makebox[0pt]{\hskip#2in
\includegraphics[width=#3\textwidth]{#1}}\vskip#4pt\hfill}

\documentclass{emulateapj}
\usepackage{amsmath,amssymb,graphicx}

\newcommand{\Ob}{\Omega_b}
\newcommand{\Om}{\Omega_m}
\newcommand{\Ol}{\Omega_\Lambda}
\newcommand{\Mpch}{h^{-1}{\rm Mpc}}
\newcommand{\kpch}{h^{-1}{\rm kpc}}
\newcommand{\Msunh}{h^{-1}M_\odot}
\newcommand{\zreion}{z_{\rm reion}}

\shorttitle{Imprint of Inhomogeneous Hydrogen Reionization}
\shortauthors{Cen, McDonald, Trac, \& Loeb}

\begin{document}

\title{Probing the Epoch of Reionization with the Lyman Alpha Forest at 
$z\sim 4-5$}

\author{Renyue Cen$^1$, Patrick McDonald$^2$, Hy Trac$^3$, \& Abraham Loeb$^3$} 

\affil{
$^1$ {\it Department of Astrophysical Sciences, Princeton
University, Princeton, NJ 08544, USA}
$^2$ {\it Canadian Institute for Theoretical Astrophysics, University of Toronto, Toronto, ON M5S 3H8, Canada}
$^3$ {\it Harvard-Smithsonian Center for Astrophysics, Cambridge, MA 02138, USA}\\ 
}

\begin{abstract}

The inhomogeneous cosmological reionization process leaves tangible 
imprints in the intergalactic medium down to $z\sim 4-5$.
The Lyman-$\alpha$ forest flux power spectrum provides
a potentially powerful probe of the epoch of reionization.
With the existing SDSS I/II quasar sample 
we show that two cosmological reionization scenarios, 
one completing reionization at $z=6$ and the other at $z=9$,
can be distinguished at $\sim 7\sigma$ level by
utilizing \lyaf\ absorption spectra at $z=4.5\pm 0.5$, 
in the absence of other physical processes that may also affect the $\lya$ flux power spectrum.
The redshift range $z=4-5$ may provide the best window,
because there is still enough transmitted flux and quasars to measure precise 
statistics of the flux fluctuations,
and the IGM still retains a significant amount of memory of reionization.

\end{abstract}

\section{Introduction}

The history of cosmological reionization is presently primarily
constrained by the cosmic microwave background observations
of WMAP (Wilkinson Microwave Anisotropy Probe) \citep[][]{2009Dunkley}
and the SDSS (Sloan Digital Sky Survey) quasar absorption spectra.
The former gives an integral constraint, strongly
suggesting that cosmological reionization may well be underway at $z\sim 12$,
while the latter provides a solid anchor point 
at $z\sim 6$ when the universe became largely transparent to Lyman limit photons
\citep[e.g.,][]{2001Fan, 2001Becker, 2002CenMcDonald, 2006Fan}. 
At $z\ge 6.3$ the lower bound on the neutral hydrogen fraction, $x$,
of the IGM provided by SDSS observations
is, however, fairly loose at $x \ge 0.01$.
Thus, exactly when most of the neutral hydrogen became reionized
is yet unknown and there are many
possible scenarios that could meet the
current observational constraints
\citep[e.g.,][]{2001BarkanaLoeb, 2003Cenb, 2003HaimanHolder, 2006FCK, 2007WyitheCen, 2007Becker}. 

The process of inhomogeneous cosmological reionization 
leaves quantifiable and significant imprints 
on the thermal evolution of the IGM \citep[][]{2008TracCenLoeb}. 
In this {\it Letter}, we show that the \lyaf\ flux
spectrum at moderate redshift
$z=4.5\pm 0.5$ sensitively depends on and hence provides a very 
powerful probe of the epoch of reionization.

\section{Reionization Models}

We use a hybrid code to accurately compute the reionization process,
which consists of a high-resolution N-body code,
a shock-capturing TVD hydro code and a ray-tracing
radiative transfer (of Lyman limit photons) code.
The reader is referred to \citet{2008TracCenLoeb} for more details.
We use the best fit WMAP 5-year cosmological parameters: 
$\Om=0.28$, $\Ol=0.72$, $\Ob=0.046$, $h=0.70$, $\sigma_8=0.82$, and $n_s=0.96$ 
\citep{2009Komatsu}.
We use $29$ billion dark matter particles on an effective mesh with $11,520^3$ cells 
in a comoving box of $100\ \Mpch$, yielding a  
particle mass resolution of $2.68\times10^6\ \Msunh$ allowing 
us to resolve all atomic cooling dark matter halos.
A total of $N=1536^3$ gas cells of size $65$kpc/h are used and we trace  
five frequency bins at $>13.6$~eV with the ray-tracing code.
The star formation rate is controlled by the halo formation history.
We adjust the ionizing photon escape fraction to arrive at two models, 
where reionization is completed early ($z\sim9$) and late ($z\sim6$), respectively;
note that the halo formation histories in the two models are identical.

\section{Results}

\begin{figure*}[t]
\center
%\hspace{-0.3in}\includegraphics[width=7.4in]{fig1.pdf}
\hspace{-0.3in}\includegraphics[width=7.4in]{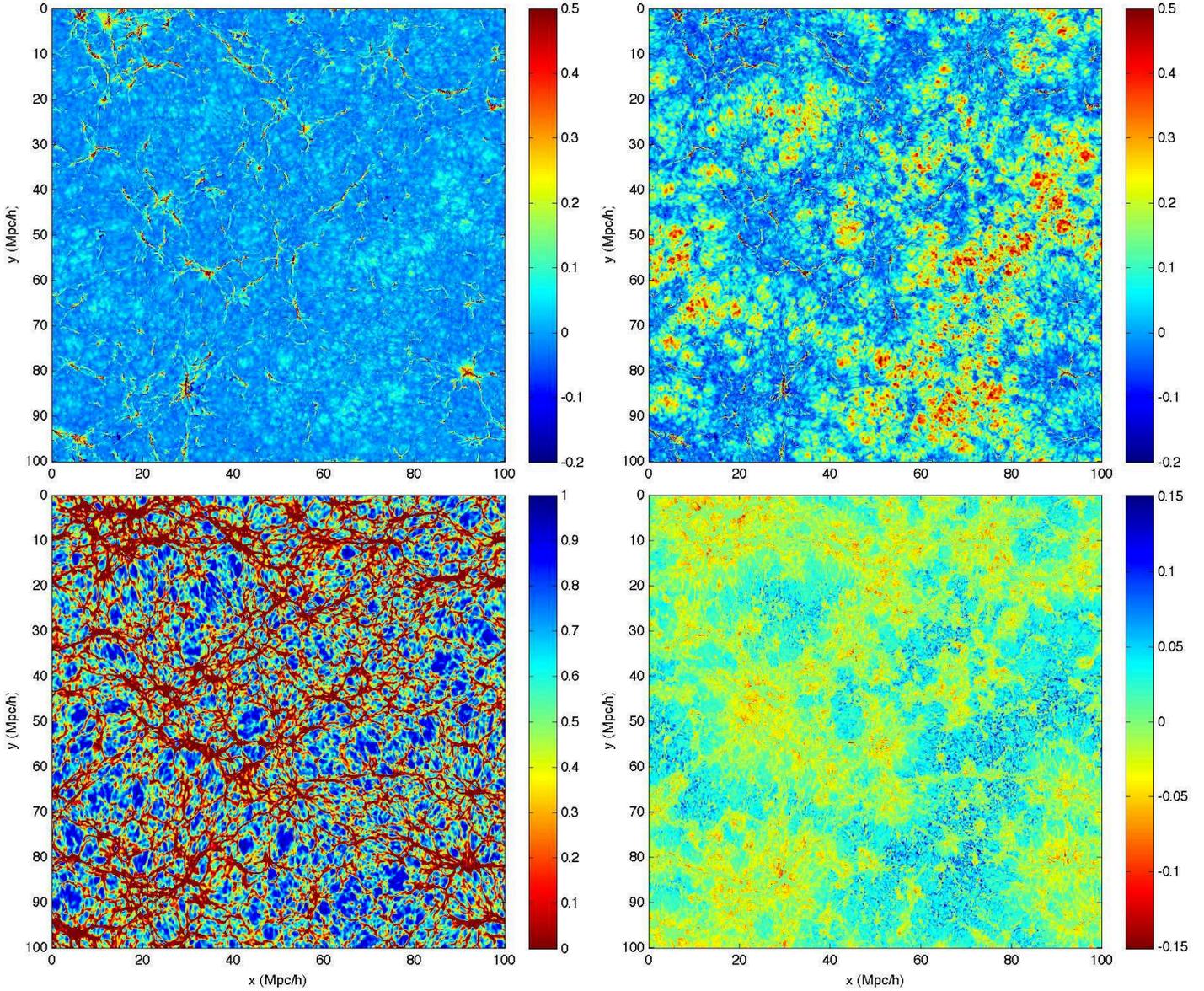}
\caption{
Top panels show the log of the ratio of gas temperature 
from the simulation to that prescribed by a fixed EoS at $z=4$,
for the early (left) and late (right) reionization model, respectively. 
We use EoS formula $T=T_0(\rho/\rho_0)^{0.62}$, where $T_0$ is the temperature
at mean density $\rho_0$ in each model.
The slice shown has a size $(100\ \Mpch)^2$ with a thickness equal to two hydro cells (130 $\kpch$).
The distribution of flux transmission, F(early)$=\exp(-\tau({\rm early}))$,
for the late reionization model is shown in the bottom left.
The flux difference between the two models:
F(late)$-$F(early)$=\exp(-\tau({\rm late}))-\exp(-\tau({\rm early}))$
is shown in the bottom right panel.
}
\label{fig:Tmaps}
\end{figure*}

Previous studies \citep[e.g.,][]{2004FZH, 2006Iliev, 2008Lee} have shown that 
the reionization process proceeds in an inside-out
fashion, where regions around high density peaks get reionized first.
H II regions initially surround isolated galaxies that formed in high density peaks.
With time these H II regions expand and lower density (void) regions 
are eventually engulfed by the expanding H II regions stemming from high density peaks. 
Consequently, the redshift of reionization of each individual spatial point,
$\zreion$, is highly correlated with the underlying large-scale density field,
with the positive correlation extending down to scales $\sim 1\ \Mpch$,
as we have shown earlier \citep[][]{2008TracCenLoeb}.
Once an expanding region is photo-ionized and photo-heated,
it would cool subsequently due to adiabatic expansion
and other cooling processes (primarily Compton cooling at high redshift),
countered by photoheating of residual recombining hydrogen
atoms (on the time scale of recombination) \citep[e.g.,][]{2002Theuns, 2003HuiHaiman}.
As a result, the strong correlation between 
$\zreion$ and the underlying large-scale density 
is manifested in a strong anti-correlation between 
the temperature and the underlying large-scale density field.
Specifically, different regions 
of the same low densities $\delta \le $ a few
(without large-scale smoothing in this case)
would display a large, long-range-correlated, dispersion in temperature, 
immediately following the completion of reionization
\citep[e.g.,][]{2008TracCenLoeb}.
(Note that virialized regions are not affected and do not retain
any information of reionization in this regard.)

Both the anti-correlation between temperature 
and the underlying large-scale density 
and the consequent temperature dispersion at a fixed density
weaken as time progresses
and the temperature-density relation asymptotically approaches a so-called
equation-of-state (EoS), 
a one-to-one mapping from IGM density to temperature \citep[][]{1997HuiGnedin},
with $T=T_0(\rho/\rho_0)^{0.62}$ in the late-time limit.
However, at the redshift range $z=4.5 \pm 0.5$, the IGM 
has not had enough time to have completely relaxed
to this state prescribed by the EoS such that quantitatively
significant deviations from a deterministic EoS exist, if the universe was 
reionized, say, 
at $z_{ri}\sim 6-8$.
The deviations from a simple temperature-density relation 
are larger for smaller $z_{ri}$ at a given
observed redshift.

In Fig.~\ref{fig:Tmaps} we show 
the log of the ratio of gas temperature 
from the simulation to that prescribed by the asymptotic EoS at $z=4$
in a slice of size $(100\ \Mpch)^2$ with a thickness equal to two hydro cells (130 $\kpch$), 
for the early (top left panel) and late (top right panel) reionization model, respectively. 
The fields have been smoothed on cells of comoving length $130\ \kpch$. 
The small reddish/yellowish regions seen in the top left
panel correspond to virialized regions, for which 
the plotted ratio does not contain useful information.
But these regions show clearly the location of ionizing sources.
We see striking differences in temperature distributions
between the two reionization models with respect to their respective
asymptotic EoS values.
In the early reionization model (top left panel)
most of the regions have blue color (i.e., the ratio equal to
$\sim 1$) and appear to have mostly relaxed to the state predicted by the
asymptotic EoS, 
while some low density regions in the voids still display yellowish color
with a temperature that is higher than that of the asymptotic EoS by $30-50\%$.
On the other hand, in the late reionization simulation (top right panel),
while regions just outside the shock-heated filaments and halos (bluish color)
have largely relaxed to the asymptotic EoS, 
regions of comparable local densities in the voids
are much hotter than that of the asymptotic EoS, by a factor of $1.5-2.5$.

Because the neutral hydrogen fraction in regions of moderate density is 
determined
by the balance between photoionization rate and recombination rate, 
the latter of which is a function of temperature,
the two different temperature distributions in the two reionization models
result in different large-scale neutral hydrogen distribution.
In the bottom left panel of Fig ~\ref{fig:Tmaps}
we show the expected flux transmission,
F(early)$=\exp(-\tau({\rm early}))$, 
for the early reionization model,
where $\tau({\rm early})$ is the $\lya$ optical depth computed
based on the distribution of neutral hydrogen density, gas peculiar velocity
and temperature at $z=4$ in the early reionization model.
In computing the neutral hydrogen fraction we have used
a uniform background radiation field with its amplitude adjusted
such that both models yield the same mean transmitted flux of $<F>=0.43$ 
at $z=4$, as observed \citep[][]{2006Fan}.
In the bottom right panel the flux difference between the two models,
F(late)$-$F(early), is shown,
%We do not use the actual radiation field from the ray-tracing code
%in the simulations for two reasons.
%First, a significant portion of the fluctuating radiation background 
%\citep[][]{2004MeiksinWhite, 2004Croft, 2005McDonald} is not included
%due to lack of treatment of contributions from quasars in the simulations.
%Second, relevant $\lya$ sink terms including Lyman limit systems
%that are below our resolution are no longer properly treated,
%once the universe has mostly reionized.
%Third, small-scale, cell-to-cell fluctuations in the radiation field
%due to ray merging of sometime a small number of rays in some regions in the ray-tracing code,
%while have little effect on the large-scale temperature distribution,
%produce visually unphysical features.
where it is clearly seen 
that the transmitted $\lya$ flux 
is significantly affected by the temperature difference at $z=4$,
resulting in fractional difference in the transmitted flux
in the voids between the two models of $\sim 15\%$ (blue regions).
Specifically, there is more transmitted flux in the void regions in the late
reionization model, compensated by comparably reduced transmitted flux in high density regions.
It is noted that, at $z\ge 4$,
the majority of transmitted $\lya$ flux
comes from the lowest density regions of $\delta \le $ a few.

\begin{figure}[t]
\center
\includegraphics[width=3.2in]{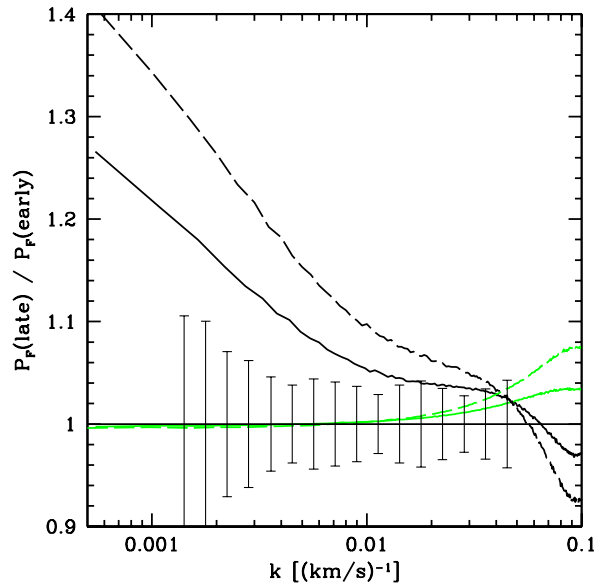}
%\vskip -0.75in
\caption{
Black solid and dashed curves are the ratio of flux power spectrum in 
the late reionization model to that in the early reionization model
at $z=4$ and $z=5$, respectively.  
Also shown as the two green curves are the corresponding ratios 
produced by replacing 
the real temperature in each simulation by that prescribed
by the EoS given density (the same EoS in both simulations).
The black error bars are the error one can expect from the full SDSS I/II 
sample plus existing high resolution data.
The error bars will be approximately uncorrelated.
A formal analysis of the $16$ data points indicates
that the two reionization models can be differentiated at $7\sigma$ level.
}
\label{fig:Pk}
\end{figure}

Fig.~\ref{fig:Pk} shows
the ratio of flux power spectrum in 
the late reionization model to that in the early reionization model
at $z=4$ (black solid) and $z=5$ (black dashed). 
It appears that the large-scale anti-correlation between density and deviations
from a single EoS in the late reionization model leads to a 
significant amount of extra power in the flux spectrum (specifically, 
relatively high
temperatures in late-reionizing under-dense regions lead them to produce even
less absorption than they otherwise would).
The difference between the flux power spectra of the two reionization models
increases with scale, reaching $20\%$ at $k=0.001$(km/s)$^{-1}$ at $z=4$;
the difference is still larger at $z=5$ ($\sim 30\%$), as expected, due to 
still larger difference in the temperature hence flux transmission
between the two reionization models.
The black error bars indicate the statistical 
errors expected with the full SDSS I/II sample (completed, but not yet 
fully analyzed), 
With $z\sim 4$ SDSS I/II data plus existing high resolution data,
one can distinguish formally 
between these two reionization models at $7\sigma$ level.
However, we note that the statistical differences between the two models
are un-marginalized, i.e., not taking into account other physical effects that
affect the $\lya$ flux power spectrum determination \citep[e.g,][]{2005McDonald}. 
Therefore, the quoted statistical significance only serves as an 
indication of the potential power of this statistics.

For comparison, the corresponding flux power spectrum ratios
at $z=4$ and $z=5$, if both models follow the same EoS (given density),
are shown (green curves) in Fig. ~\ref{fig:Pk}.
In this case, aside from the relatively small difference on 
small scales due to cumulative dynamical affects on the gas density
by the difference in the gas pressure histories,
the two models have identical flux power spectrum on large scales.
This clearly demonstrates that the large difference in the flux power spectra
between the two reionization models (black curves in Fig. ~\ref{fig:Pk}) 
is a result of large differences in the contemporaneous temperature distributions.

\section{Discussion}

This effect of inhomogeneous reionization on the flux power spectrum 
was explored earlier by \citet{2006Lai} at $z=3$,
based on a semi-analytic model.
Their focus is on $z=3$ and 
found that, on large scales, $k\sim 0.001$(km/s)$^{-1}$, 
temperature fluctuations lead to an increase in the $z\sim 3$ 
flux power spectrum by at most 10\%.
Our focus here is at higher redshifts $z=4-5$
and the effects, not surprisingly, are larger and potentially more discriminating.
%Given the difference in the modeling and their lower redshift
%of study, compared to ours ($z=4-5$),
%it seems that our results are not in disagreement with theirs.
%We note that reionization process is complex and it is not
%surprising that simple semi-analytic methods
%produce results that are quantitatively different from detailed radiative transfer
%simulations.

A fluctuating radiation background, produced largely by radiation from sparsely
distributed quasars but also by galaxies, 
can affect the flux power spectrum 
\citep[][]{2004MeiksinWhite, 2004Croft, 2005McDonald}. 
Larger fluctuations in the radiation background
give rise to larger amplitudes of the flux power spectrum
at large scales \citep[e.g.][Figures 6,7 therein]{2005McDonald}. 
This enhancement of the flux power spectrum on large scales
due to a fluctuating radiation background 
will be in addition to what is caused by the gas temperature fluctuations shown here,
if QSOs were dominant.
The radiation contribution from stars may be more dominant at
the redshift range of concern here \citep[e.g.,][]{2009Faucher-Giguere}.
Star formation is known to be biased and 
hence higher density regions, on average, tend to have higher radiation field
than lower density regions.
Thus, the two effects due to a fluctuating radiation background and 
an inhomogeneous reionization process may be partially degenerate or have
a tendency to cancel each other's contribution,
although there is a possibility that the radiation fluctuations may be relatively modest
\citep[e.g.,][]{2009Mesinger}. 
A more careful modeling of the contribution from quasars as well as
radiation sinks (such as Lyman limit systems) is required in a comprehensive modeling.
The purpose of this {\it Letter} is to demonstrate that,
if the effects on the $\lya$ flux power spectrum determination
due to the epoch of reionization were the only relevant ones,
then a precise measure of the flux power spectrum with the full SDSS I/II
data will be able to place a very tight constraint
on the epoch of reionization.

However, a detailed comparison between models and SDSS I/II observations
requires a full analysis of all
astrophysical/cosmological processes that may affect 
the determination of the flux power spectrum 
and some of them may be degenerate to varying degrees \citep[][]{2005McDonald}, 
including fluctuating radiation field, damped $\lya$ systems,
galaxy formation feedback, initial photoheating temperature (i.e., related to 
IMF of high redshift galaxies),  X-ray heating, 
He II reionization, among others,
before its statistical potential can be precisely marginalized and quantified.
We will perform such an analysis in a future study.

%\newpage
\section{Conclusions}

Utilizing state-of-the-art radiative transfer 
hydrodynamic simulations of cosmological reionization,
we put forth the point that the inhomogeneous reionization process
imprints important and quantitatively significant signatures in the 
intergalactic medium at $z=4.5 \pm 0.5$ that 
can be probed by the $\lya$ forest in the quasar absorption spectra.
We illustrate that with $\lya$ forest data at $z=4-5$
to be provided by the SDSS I/II full data sample,
one may be able to distinguish between two cosmological epochs of reionization,
one at $z=6$ and the other $z=9$ at $7\sigma$ level,
if they were the only effects on the determination of the $\lya$ flux power spectrum.

\acknowledgments
We thank J.~Chang at NASA for invaluable supercomputing support. 
This work is supported in part by NASA grants NNG06GI09G and NNX08AH31G.
Computing resources were in part provided by the NASA High-End Computing (HEC) 
Program through the NASA Advanced Supercomputing (NAS) Division at Ames 
Research Center.  PM acknowledges support
of the Beatrice D. Tremaine Fellowship.
HT is supported by an Institute for Theory and Computation Fellowship.

\bibliographystyle{apj}
\bibliography{astro}

\end{document}